\documentstyle[pra,aps]{revtex}
\begin{document}
\draft
\title{Disorder-induced solitons in conjugated polymers}
\author{Maxim V. Mostovoy\cite{Perm}, Marc Thilo Figge, and Jasper~Knoester}
\address{Institute for Theoretical Physics, Materials Science Center\\
University of Groningen, Nijenborgh 4, 9747 AG Groningen, The Netherlands}
\date{\today}
\maketitle
\begin{abstract}
\widetext
\leftskip 54.8pt
\rightskip 54.8pt
We show that weak off-diagonal disorder in degenerate ground
state conjugated polymers results in a finite density of randomly
positioned kinks (solitons and antisolitons) in the lattice
dimerization.  For realistic values of the disorder, these kinks
should clearly show up in the optical and magnetic properties.
\end{abstract}

For over thirty years, conjugated polymers have provided a rich
field of fundamental and technological research \cite{Kies}.
They belong to a large class of quasi-one-dimensional Peierls
materials, in which the lattice distorts due to its interaction
with itinerant electrons \cite{PhysTod}.  Much attention has been
devoted to trans-polyacetylene, which has a doubly degenerate
ground state.  The degeneracy allows for topological excitations,
solitons, which are kinks in the lattice dimerization accompanied
by a local distortion of the electron density \cite{SSH}.
Solitons provide a simple explanation of the unusual properties
of doped trans-polyacetylene \cite{Kies,SSH}.  They also feature
in quantum lattice fluctuations, which lead to subgap optical
absorption \cite{AK} and may be responsible for an enhanced
nonlinear optical response \cite{EL}.  Solitons were first found
in the free-electron SSH model \cite{SSH,SU}.  Their topological
nature, however, allows them to survive also in the presence of
on-site Coulomb repulsion (Peierls-Hubbard model) \cite{Cam}.

Although real polymer materials contain many types of defects,
most theoretical studies have been restricted to perfect chains.
Due to the quasi-one-dimensional nature of polymers, however, any
disorder will play a major role in determining their properties.
In this Letter, we study the effects of weak disorder in the
electron hopping amplitudes in polyacetylene chains.  Such
(off-diagonal) disorder arises from random chain twists that
diminish the overlap between the $\pi$-orbitals of neighboring
carbon atoms.  In previous analytical studies, off-diagonal
disorder was modeled by small fluctuations of the Peierls order
parameter around an average value \cite{OE,Kim}.  Numerical
calculations showed, however, that if the lattice configuration
of a weakly disordered chain described by the SSH model is
allowed to relax, it may contain solitons, {\em i.e.}, large
(nonperturbative) variations of the order parameter
\cite{Iwan,Wada}.  This implies that the lattice response to the
disorder cannot be neglected.  An analytical theory for this
phenomenon and an estimate (either numerical or analytical) for
the density of thus created kinks, have sofar not been given.  In
this Letter, we address these issues and we show that, in fact,
arbitrarily small off-diagonal disorder leads to a finite density
of solitons.  Our arguments also hold in the presence of electron
correlations.

We consider a half-filled chain, in which itinerant ($\pi$)
electrons interact with the classical carbon lattice.  We assume
that the chain Hamiltonian contains three contributions: (i) the
electron hopping between adjacent carbon atoms, (ii) the elastic
energy of the lattice, and (iii) the Coulomb repulsion between
the electrons.  The hopping amplitude has the form,
\begin{equation}
\label{hop}
t_{m,m+1} = t_0 + \alpha (u_m - u_{m+1}) + \delta t_{m,m+1} \;\;.
\end{equation}
Here the first term is the bare amplitude and the second term
describes the electron-phonon interaction, where $\alpha$ is the
coupling constant and $u_m$ denotes the displacement of the
$m$'th carbon atom from its uniform-lattice position.  Finally,
the third term is a random contribution, which we include to
describe the conformational disorder.  We assume that the
fluctuations $\delta t_{m,m+1}$ are frozen (``quenched''
disorder) and that they are independent for different links.  We
will restrict ourselves to zero-temperature properties.

As we shall shortly see, for weak disorder the spatial extension
of a disorder fluctuation necessary to create a kink is much
larger than one lattice unit, which allows us to use a continuum
model.  Then Eq.(\ref{hop}) is replaced by,
\begin{equation}
\label{model}
\Delta(x) = \Delta_{lat}(x) + \eta(x)  \;,
\end{equation}
where the lattice dimerization $\Delta_{lat}(2ma) = \alpha
(u_{2m-1}-2u_{2m}+u_{2m+1})$
($a$ is the lattice constant) describes the alternating part of
the hopping amplitude determined by the shifts $u_m$ of the
carbon atoms (cf.  the continuum version of the SSH model
\cite{TLM}) and $\eta(2ma) = \delta t_{2m-1,2m} - \delta
t_{2m,2m+1}$ describes the disorder part with a Gaussian
correlator,
\begin{equation}
\label{Gauss}
\langle \eta (x) \eta(y) \rangle =
A \delta(x - y) \;.
\end{equation}
Note, that while chain twists always decrease the hopping
amplitudes ($\delta t_{m,m+1} < 0$), $\eta(x)$ can be both
positive and negative, as it is the alternating part of the
fluctuations.

It should now be realized that in a consistent treatment of
disorder, $\Delta_{lat}(x)$ implicitly depends on the $\eta(x)$.
The reason is that at zero temperature $\Delta_{lat}(x)$ for a
particular disorder realization is found by minimizing the total
(electronic plus lattice) energy.  The total energy reads,
\begin{equation}
\label{energy}
E = E_{el}\left[ \Delta(x) \right] +
\frac{1}{\pi \lambda v_F} \int \!\!dx \Delta_{lat}^{2}(x) \;,
\end{equation}
where $\lambda = \frac{4 \alpha^2}{\pi t_0 K}$ is the
dimensionless electron-phonon coupling constant ($K$ is the
spring constant), $v_F = 2 a t_0$ is the bare value of Fermi
velocity, and we set $\hbar = 1$.  The electron energy depends on
$\Delta(x)$ defined by Eq.(\ref{model}), which includes the
changes of the hopping amplitudes due to both the lattice
distortion and the disorder.  We stress that in what follows we
will not need an explicit expression for the electron energy,
which enables us to include the electron-electron interaction.
We will assume, however, that in the absence of disorder the
total energy of a half-filled chain has two minima, both
corresponding to a uniform dimerization, $\Delta(x) = \pm
\Delta_0$.  We will also assume that the model admits soliton
solutions, {\em i.e.}, apart from the above minima, the total
energy has an infinite number of extrema
$\Delta_N(x|\mbox{\boldmath $z$})$ corresponding to the presence
of $N$ kinks (solitons and antisolitons interpolating between
$-\Delta_0$ and $+\Delta_0$ and vice versa), whose positions are
described by the $N$-dimensional vector $\mbox{\boldmath $z$} =
(z_1, z_2,\ldots,z_N)$.  Both the SSH and the Peierls-Hubbard
model have these properties.  If the average separation between
neighboring kinks is large compared to the correlation length
$\xi_0 = v_F/\Delta_0$, the configuration with $N$ kinks has
energy,
\begin{equation}
E_N = E_0 + N E_s\;,
\end{equation}
where $E_s$ is the energy needed to create a single soliton.  The
kinks can be either charged and spinless or neutral with spin
$\frac{1}{2}$ \cite{SSH}.  In the SSH model, the two types have
the same energy, while in the presence of Coulomb repulsion the
neutral soliton is energetically favorable \cite{Cam,KH} and in
undoped chains will be the only type of kink that occurs at zero
temperature.

At weak disorder, $|\eta(x)| \ll \Delta_0$, linear response
theory predicts small fluctuations of the ground state lattice
configuration around the uniformly dimerized lattice.  We
will show, however, that for arbitrarily weak disorder the ground
state configuration cannot be found using linear response theory,
as the configuration with small fluctuations around a chain
containing a number of solitons may have a lower energy
than the one with small fluctuations around uniform dimerization.

To first order in $\eta(x)$, the correction to the energy of a
multikink configuration reads,
\begin{equation}
\label{EN}
\delta E_N = - \frac{2}{\pi \lambda v_F} \int \!\!dx
\Delta_N(x|\mbox{\boldmath $z$}) \eta(x) \;,
\end{equation}
where the extremum condition for the ordered configuration
$\Delta_N(x|\mbox{\boldmath $z$})$ was used.  Hence, the change
of the total energy of the uniformly dimerized state
($\Delta(x)=\Delta_0$) due to disorder reads,
\begin{equation}
\delta E_0 = -
\frac{2 \Delta_0}{\pi \lambda v_F} \int \!\!dx \eta(x) \;.
\end{equation}
Consider now, on the other hand, the same chain with an
antisoliton at $z_1$ and a soliton at $z_2$, chosen in such a way
that the whole disorder fluctuation lies
between $z_1$ and $z_2$.
The change of energy due to disorder for this configuration
equals $- \delta E_0$, because between $z_1$ and $z_2$
$\,\Delta_2(x|z_1,z_2) \approx - \Delta_0$.  We thus find that in
the disordered chain, the configuration with a
soliton-antisoliton pair is energetically favorable to the
uniform configuration if
\begin{equation}
\label{pair}
- \int \!\!dx \eta(x) > \gamma \lambda v_F \;.
\end{equation}
Here $\gamma = \frac{\pi E_s}{2 \Delta_0}$, which in the case of
the SSH model equals 1.  It should be stressed that as long as
the kink density is indeed small compared to $1/\xi_0$, the
entire effect of electron-electron interactions is contained in
the factor $\gamma$.  We note that, no matter how small the
disorder is, the fluctuations in $\int \!\!dx \eta(x)$
grow with the chain size, so that for a sufficiently long chain
the inequality Eq.(\ref{pair}) will certainly be fulfilled.

The creation of kinks by off-diagonal disorder is illustrated in
Fig.~1.  The thick line is the ground state order parameter
$\Delta_m = t_{2m-1,2m} - t_{2m,2m+1}$ for a discrete chain of
160 carbon atoms, obtained by numerically minimizing the total
energy for one particular realization of the disorder.  In this
example, we neglected the Coulomb repulsion and used standard SSH
parameters.  It is clearly seen that the order parameter
fluctuates near a soliton-antisoliton pair configuration.  The
disorder realization that was used as input is visualized by the
thin line, which gives $\Delta_0 + \eta_m = \Delta_0 + \delta
t_{2m-1,2m} - \delta t_{2m,2m+1}$ ($\Delta_0 \approx 0.7{\mbox
eV}$).  The thin line in fact directly represents the
order parameter along the chain if we neglect the lattice
relaxation.  In that case, no lattice kinks are created.

In order to obtain an estimate for the density of
disorder-induced kinks, we consider a disorder fluctuation with
spatial size $l$ and constant value $\eta_l$.  Then, according to
Eq.(\ref{pair}), the threshold for the creation of a
soliton-antisoliton pair is,
\begin{equation}
\label{etal}
|\eta_l| = \frac{\gamma \lambda v_F}{l} \;.
\end{equation}
By requiring the probability density of such a fluctuation,
\begin{equation}
\label{p}
p =
\exp \left( - \frac{1}{2 A}
\int\!\!dx \eta(x)^2 \right) =
\exp \left( - \frac{(\gamma \lambda v_F)^2}{2 A l} \right) \;,
\end{equation}
to be of the order of unity, we obtain the average density of
kinks:
\begin{equation}
\label{answer}
n = \frac{1}{l} = \frac{A}{\left(
\gamma \lambda v_F \right)^2} \;.
\end{equation}
The condition $p \sim 1$ ensures that the typical value of
$\int_x^{x+l}\!\!dx^{\prime} \eta(x^{\prime})$ is large enough to
stabilize a soliton-antisoliton pair.  The probability to find
such a pair with separation much smaller than the average given
by Eq.(\ref{answer}), is suppressed (cf.~Eq.(\ref{p})) as it
requires a large disorder fluctuation.  On the other hand,
lattice configurations in which the average distance between the
kinks is significantly larger than $l$ given by Eq.(\ref{answer})
are statistically less likely.  Similar arguments were used to
estimate the typical domain size for the Ising model in a random
magnetic field \cite{IM}.  The analogy is not accidental: it can
be shown that the statistical properties of the disorder-induced
kinks are indeed described by the latter model.

We next address the question whether our first-order expansion
around multi-kink configurations is a meaningful approach.
First, we estimate second-order effects.  To this end, we
again consider the fluctuation with spatial size $l$ and
magnitude $|\eta_l|$.  In this case, the ratio ${\delta^{(2)}E}/
{\delta^{(1)}E}$ of the second- and first-order corrections to
the total energy can be expressed in terms of the renormalized
frequency of the phonon with wave-vector $\pi / a$.  This
frequency can be calculated in the two limiting cases for the
on-site electron-electron interaction $U$.  We then find:
\begin{equation}
\frac{\delta^{(2)}E}{\delta^{(1)}E} =
\frac{\xi_0}{l} \left\{
\begin{array}{rl}
\frac{1}{2}(1 - \lambda), & \mbox{ for} \;\;U = 0 \\
\frac{1}{4}\gamma \lambda, & \mbox{ for} \;\;U \gg t_0
\end{array} \right.
\end{equation}
plus higher order terms in powers of $\xi_0 / l$.  Therefore, our
approach is valid as long as the average distance between the
kinks is much larger than the correlation length (dilute gas of
kinks).

Second, we should stress that, although the energies of the
multi-kink configurations are calculated perturbatively, the
stabilization of kinks by weak disorder is clearly a
non-perturbative effect.  The reason is that for growing disorder
the minimal-energy lattice configuration with a finite density of
kinks does not evolve continuously from the uniformly dimerized
configuration.  In fact, as long as the second-order correction
to the energy due to disorder is relatively small, each multikink
lattice configuration does not change much in the presence of
weak disorder.  Nevertheless, their energies, even in the linear
regime, may change appreciably for sufficiently long chains.
This fact is illustrated in Fig.~2, where we plot the energies of
several multi-kink configurations, with 0 to 4
soliton-antisoliton pairs, as a function of the magnitude of the
disorder.  Here, the shape of the disorder fluctuation is assumed
to be fixed, and the slopes of the plotted lines depend on the
particular choice of the kink positions.  From this figure it is
clear that the increase of the number of soliton-antisoliton
pairs in the minimal-energy configuration for a given shape of
the disorder fluctuation occurs discontinously in the region
where the energies are still linear in the disorder.

Elsewhere, we show in detail that the estimate for the density of
disorder-induced kinks Eq.(\ref{answer}) is, in fact, the exact
result in the limit where the average distance between kinks is
much larger than the correlation length $\xi_0$ (dilute gas of
kinks) and much smaller than the length of the chain $L$.  The
exact solution was obtained by noticing that the disorder
averaged free energy of the chain at sufficiently low temperature
has the form of a matrix element of the Green function describing
the relaxation of a spin-$\frac{1}{2}$ in a random magnetic
field.  The coordinate along the chain plays the role of the
(imaginary) time in which the relaxation takes place, while the
kinks correspond to spin-flips.  The Green function was found by
solving the corresponding Fokker-Planck equation.

We now discuss the consequences of the disorder-induced solitons
for the optical and magnetic properties.  In the absence of
electron-electron interactions, the finite density of kinks leads
to the appearance of a peak in the single-particle density of
states at $\varepsilon = 0$, because each kink carries an
electron state with zero energy \cite{SSH}.  At weak disorder
this peak is sharp since the distances between the kinks are
large, so that the splitting of the energies of the electron
states localized near different kinks due to their mutual
interaction is very small ($\sim \exp \left(- \frac{l}{\xi_0}
\right)$).  The peak in the density of states should result in a
midgap peak in the optical absorption spectrum.  In undoped
trans-polyacetylene, however, this peak is not observed.  This
can be ascribed to a small magnitude of disorder, which is,
however, unlikely, since the average (presumably disorder
limited) conjugation length of trans-polyacetylene is known to be
of the order of several tens of carbon atoms \cite{Silbey}.

The on-site Coulomb repulsion decreases the neutral soliton
energy\cite{Cam,KH} ($\gamma < 1$), thus increasing the kink
density (Eq.(\ref{answer})).  However, at the same time, the
Coulomb interaction shifts the absorption peak resulting from the
neutral kinks towards higher energies, where it may merge with
the peak resulting from interband transitions.  This may explain
the absence of a clear midgap peak in the absorption spectrum,
but it does not reconcile the theory with magnetic measurements.
The neutral disorder-induced kinks have spin $1/2$ and should
contribute to the magnetic susceptibility.  However, the observed
Curie susceptibility for undoped trans-polyacetylene corresponds
to only one free spin per 3000 carbon atoms \cite{SII}, which is
much smaller than the average density of kinks given by
Eq.(\ref{answer}) for the disorder magnitude used in \cite{Kim}
to fit the absorption spectrum.

This contradiction between theory and experiment clearly poses
questions as to the applicability of the SSH or the
Peierls-Hubbard models to trans-polyacetylene.  Here we note
several factors that may solve this problem.  First, interchain
interactions counteract the formation of free kinks, binding them
into pairs.  Second, quantum lattice motion induces an effective
interaction between the spins of neighbouring solitons, resulting
in a spin-liquid with a singlet ground state.  Finally, the free
spins of solitons can be bound by chemical impurities.  Further
research is needed to see whether the quantitative effects of
these factors are strong enough to account for the strong
reduction in the number of observed free spins.

In summary, we have shown that in Peierls chains arbitrarily weak
off-diagonal disorder induces a finite density of neutral
solitons.  As we discussed, these solitons strongly affect the
observable properties of the chains.  We believe, that they also
play an important role in the phase transition of disordered
materials into the Peierls state.

This work is supported by the "Stichting Scheikundig Onderzoek in
Nederland (SON)" and the "Stichting voor Fundamenteel Onderzoek
der Materie (FOM)".

\section*{Figure Captions}

FIG.~1.  Numerically obtained order parameter along a
polyacetylene chain of 80 pairs of carbon atoms for one
particular (but typical) realization of off-diagonal disorder.
The thin line does not account for lattice relaxation, the
thick line does.

FIG.~2.  Illustration of the energies of several multi-kink
configurations as a function of the magnitude of the disorder
fluctuation.  From bottom to top (at zero disorder), the curves
correspond to chains with 0 to 4 soliton-antisoliton pairs.

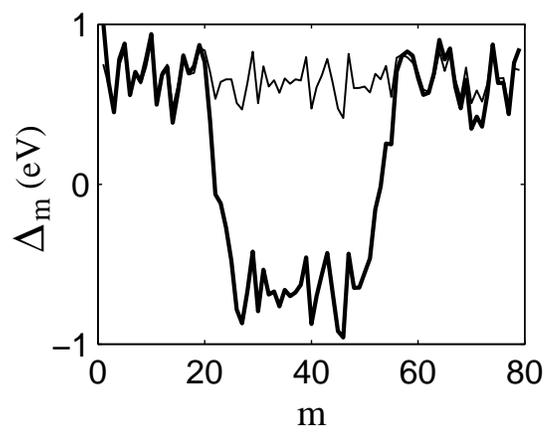
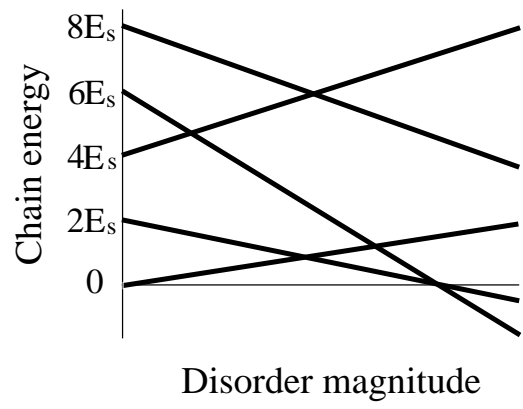
\end{document}